\documentclass[a4paper,12pt]{article}

\usepackage[latin1]{inputenc}
\usepackage{amsmath}
\usepackage{amssymb}
\usepackage[dvips]{graphicx}

\DeclareMathOperator{\Tr}{Tr}
\newcommand{\be}{\begin{equation}}
\newcommand{\ee}{\end{equation}}

\begin{document}

\begin{titlepage}
\begin{flushright}
HD-THEP-04-53\\
CERN-PH-TH-2004-260
\end{flushright}
\vspace{0.6cm}
\begin{center}
{\Large \bf \mbox{Worldlines on Orbifolds and the Fayet-Iliopoulos Term}} 
\end{center}
\vspace{0.5cm}

\begin{center}
{\large
Felix Brümmer$^{a1}$,
Michael G. Schmidt$^{a2}$,
Zurab Tavartkiladze$^{b3}$}

\vspace{0.3cm}
{\em
${}^a$ Institut für Theoretische Physik,
Universität Heidelberg\\
Philosophenweg 16, 69120 Heidelberg, Germany}\\
{\em
${}^b$ Physics Department, Theory Division, CERN, CH-1211 Geneva 23,
 Switzerland}
\end{center}
\vspace{0.4cm}
\begin{abstract}
\noindent We adapt ``string-inspired'' worldline techniques to one-loop calculations on orbifolds, in particular on the $S^1/Z_2$ orbifold. Our method also allows for the treatment of brane-localized terms, or bulk-brane couplings.  For demonstration, we reproduce the well-known result for the one-loop induced Fayet-Iliopoulos term in rigidly supersymmetric Abelian gauge theory, and generalize it to the case where soft supersymmetry breaking mass terms for the bulk scalar fields are present on the branes.
\end{abstract}
\vspace{9cm}
\footnoterule

{\small
\noindent$^1 $E-mail address: f.bruemmer@thphys.uni-heidelberg.de\\
 \noindent$^2 $E-mail address: m.g.schmidt@thphys.uni-heidelberg.de\\
\noindent$^3 $E-mail address: zurab.tavartkiladze@cern.ch}

\end{titlepage}

\section{Introduction}
There are several approaches to perturbative calculations in orbifold field theory: Decomposition of the fields into Kaluza-Klein modes, which can then be treated within ordinary four-dimensional perturbation theory, or using a mixed position-momentum space formalism \cite{mixedprops} are probably the most common ones. In this paper we outline an alternative method, relying on the ``string-inspired'' worldline formalism, which is well established in quantum field theory in four noncompact dimensions nowadays. Using worldline methods instead of standard second-quantized perturbation theory may lead to a significant reduction of calculational work in some cases \cite{ginfl}. With worldlines on orbifolds, it is also straightforward to take into account brane couplings to bulk fields in the Lagrangian, as we will see.

The worldline method is a convenient tool in quantum field theory, in particular for calculating one-loop amplitudes and effective actions. Recent interest in it arose with first-quantized string perturbation theory, where perturbative calculations are carried out by path integration over worldsheets; in the limit of infinite string tension, these become the worldlines of point particles, so one can do point particle perturbation theory in a similar manner \cite{Bern:1991aq}. This results in a significant simplification of certain one-loop calculations. Subsequently the method was reformulated and rederived by simpler means to involve first-quantized worldline path integrals \cite{Strassler:1992zr,Schmidt:1993rk}; this is of course related to Feynman's and Schwinger's old approach to propagators and loops via relativistic quantum mechanics. Recently significant progress has been made in that field: Generalizations involving multiloop calculations \cite{Schmidt:1994zj} and various types of couplings \cite{mnss,dhg} have been worked out, and worldline methods have been applied to several kinds of problems \cite{worldlineappl}.

This work is organized as follows: In section 2 we give a short reminder of the worldline one-loop effective action for a scalar field in noncompact space. In section 3 we describe how to implement compactification and orbifold boundary conditions. Section 4 contains as a simple example the calculation of the one-loop Fayet-Iliopoulos term in SUSY QED on the $S^1/Z_2$ orbifold. We demonstrate how to treat bulk-brane couplings in section 5, where we compute the one-loop renormalization of the Fayet-Iliopoulos term in the same model, but with a soft SUSY breaking scalar mass located on one of the branes. 
It is shown that such a term induces  divergences similar to those 
which would have emerged in presence of orbifold parity violating bulk 
mass operators. Finally, we draw some conclusions.

\section{Worldlines in noncompact space}

Consider a complex scalar particle of mass $m$ in $D$-dimensional noncompact Euclidean space with a self-interaction potential $V$. The classical action is
\be S[\phi,\phi^*]=\int d^Dx\,\left(|\partial_\mu\phi|^2-m^2|\phi|^2-V(|\phi|)\right) \ee
The one-loop effective action can be written as
\be \Gamma=-\Tr\log\frac{-\partial^2+m^2+V''(|\phi|)}{-\partial^2+m^2} \ee
which is written using the Schwinger proper time representation as
\be\label{schwgamma}\Gamma=\int_0^\infty\frac{dT}{T}\Tr\,\exp\left(-T\left(-\partial^2+m^2+V''(|\phi|)\right)\right)\ee
The functional trace is performed in $x$-space to give
\be\label{ftrace}\Gamma=\int_0^\infty\frac{dT}{T}\int d^Dx\,\Bigl< x\left|\exp\Bigl(-T\bigl(-\partial^2+m^2+V''(|\phi(x)|)\bigr)\Bigr)\right|x\Bigr> \ee
This expression can be evaluated using first-quantized path integrals, which leads to the worldline representation of the one-loop effective action:
\be\label{sceact} \Gamma=\int_0^\infty\frac{dT}{T}\,\mspace{-18.0mu}\int\displaylimits_{x(0)=x(T)}\mspace{-18.0mu}\mathcal Dx(\tau)\,e^{-S_\mathrm{WL}[x(\tau)]} \ee
with the worldline action given by

\be S_\mathrm{WL}[x(\tau)]=\int_0^T d\tau\left(\frac{1}{4}\dot x^2+m^2+V''(|\phi(x)|)\right) \ee
Here the path integral sums all closed trajectories of length $T$ of a fictitious particle, which are parametrized by the proper time $\tau$. UV divergences show up in the $T$ integration as $T\rightarrow 0$, and may be regularized using dimensional regularization or a simple cutoff. To do calculations it is convenient to split the path integration into a base point integral over all space and an integral over closed loops $y(\tau)$ with $y(0)=y(T)=0$,
\be\label{sceacty} \Gamma=\int_0^\infty\frac{dT}{T}\,\int d^D x_0\mspace{-24.0mu}\int\displaylimits_{y(0)=y(T)=0}\mspace{-24.0mu}\mathcal Dy(\tau)\,e^{-S_\mathrm{WL}[x_0+y(\tau)]} \ee
The normalization of the free path integral is
\be\label{norm}\int\mathcal Dy(\tau)\,\exp\left(-\int_0^T d\tau\,\frac{1}{4}\dot y^2\right)=(4\pi T)^{-D/2}\ee
and to compute $\Gamma$ explicitly, one would expand the exponential in \eqref{sceacty} using the contraction rule
\be\label{contr}\langle y^\mu(\tau_1)y^\nu(\tau_2)\rangle=-g^{\mu\nu}G^B(\tau_1,\tau_2) \ee
where the bosonic worldline Green function $G^B$ (adapted to the boundary conditions) is given by
\be G^B(\tau_1,\tau_2)=|\tau_1-\tau_2|-\tau_1-\tau_2+\frac{2}{T}\tau_1\tau_2\ee 

Similar (if slightly more complicated) formulas for one-loop effective actions exist for free fermions, for scalars or fermions coupled to background gauge fields, fields with Yukawa interactions etc.

\section{Worldlines on orbifolds}

Now let space-time not be given by $\mathbb R^D$ but by $\mathbb R^D\times\mathcal O$, where $\mathcal O$ is a $d$-dimensional compact orbifold. We take $\mathcal O$ to be the quotient of a $d$-dimensional smooth compact manifold $M$ by a group $G$ with a discrete, non-free action on $M$. $M$ itself is obtained as $M\cong N/H$, where $N$ is a $d$-dimensional smooth non-compact manifold and $H$ is a group with a discrete, free action on $N$.

A field theory on $\mathbb{R}^D\times M$ is obtained from a field theory on $\mathbb{R}^D\times N$ by requiring the fields to take equal values on each $H$-orbit.\footnote{We are not considering non-trivial (Scherk-Schwarz) boundary conditions for the compactification here for simplicity. The generalization to Scherk-Schwarz compactification should nevertheless be straightforward.} An orbifold field theory is then obtained from the field theory on $\mathbb{R}^D\times M$ by permitting only field configurations that are compatible with the orbifold symmetry in the following sense: To each field $\phi$ we assign a representation of $G$ by operators $\{P^g_\phi\}_{g\in G}$ such that for fixed $g\in G$ the $P^g_\phi$ constitute a symmetry of the Lagrangian, 
\be\mathcal L[\phi_1\ldots\phi_n]=\mathcal L[P_{\phi_1}^g\phi_1\ldots P_{\phi_n}^g\phi_n]\qquad\forall\; g\in G\ee
(by writing square brackets we indicate that $\mathcal L$ depends on the $\phi_i$ and their derivatives). A field $\phi=\phi(x,z)$, where $z$ stands for coordinates on $M$, is then required to take equal values on each of the $G$-orbits up to that symmetry,
\be\phi(x,z)=P^g_\phi\phi(x,gz)\ee
This leads to a well-defined Lagrangian on $\mathbb{R}^D\times\mathcal O$, even though the individual fields are in general not single-valued functions on the fundamental domain. An arbitrary field $\chi$ living on $\mathbb{R}^D\times M$ is projected on a field with the same orbifold symmetry properties as $\phi$ by means of the projector $Q_\phi$, which acts as
\be (Q_\phi \chi)(x,z)=\frac{1}{|G|}\sum_{g\in G}\,P^g_\phi\chi(x,gz) \ee
To do worldline calculations on $\mathbb R^D\times\mathcal O$, we proceed as for $\mathbb R^D\times N$, but include in the path integral also paths that are closed only modulo compactification and orbifold symmetry. Hence a path connecting points in $\mathbb{R}^D\times N$ that are identified in $\mathbb R^D\times\mathcal O$ is regarded as closed. For the trivial case that $G=E$ contains only the identity, we can immediately write down the path integral when going from $\mathbb{R}^D\times N$ to $\mathbb R^D\times M$:
\be\label{trivpi}\int\displaylimits_{x(0)=x(T)}\mspace{-18.0mu}\mathcal D^{D+d}x(\tau)\,e^{-S_\mathrm{WL}[x]}\quad\rightarrow\quad\sum_{h\in H}\quad\int\displaylimits_{x(0)=x(T)}\mspace{-18.0mu}\mathcal D^Dx(\tau)\int\displaylimits_{z(0)=hz(T)}\mspace{-24.0mu}\mathcal D^d z(\tau)\,e^{-S_\mathrm{WL}[x,z]}\ee
Nontrivial $G$ allow for orbifold boundary conditions, which we have to take into account. This is done by taking the functional traces that appear in the derivation of the worldline one-loop effective action over the appropriate orbifold projections of eigenstates. Consider for instance the contribution to the effective action from a complex scalar field $\phi$. In \eqref{schwgamma}, we now restrict the trace to states with the same orbifold symmetry as $\phi$. Hence we have formally
\be \Gamma=\int_0^\infty\frac{dT}{T}\sum_n\left<Q_\phi\chi_n\left|\exp\left(-T\left(-\partial^2+m^2+V''(|\phi|)\right)\right)\right|Q_\phi\chi_n\right>\ee
where the $\{\chi_n\}$ form a complete orthonormal system on $\mathbb{R}^D\times M$. How exactly this is evaluated depends of course on how the $P^g_\phi$ operators act; in general, terms involving matrix elements between $G$-equivalent points will appear in \eqref{ftrace}.

In the most common examples of orbifolds in field theory, $M$ is a torus. In that case, it is possible to set $N=\mathbb R^d$, and to represent the action of $H$ on $\mathbb R^d$ as a $d$-dimensional lattice $L$. We must then replace
\be \label{toruspi}\int\displaylimits_{x(0)=x(T)}\mspace{-18.0mu}\mathcal D^{D+d}x(\tau)\,e^{-S_\mathrm{WL}[x]}\quad\rightarrow\quad\sum_{g\in G}\;c_g\;\sum_{l\in L}\quad\int\displaylimits_{x(0)=x(T)}\mspace{-18.0mu}\mathcal D^Dx(\tau)\int\displaylimits_{z(0)=g(z(T)+l)}\mspace{-36.0mu}\mathcal D^d z(\tau)\,e^{-S_\mathrm{WL}[x,z]}\ee
with the $c_g$ numbers to be determined from the orbifold symmetry properties of the field in question. Since we know how to do worldline calculations on $\mathbb R^{D+d}$, we can use the replacement prescription \eqref{toruspi} to do explicit calculations on a large class of orbifolds. We will demonstrate a particularly simple one in the following section.

\section{Example: The Fayet-Iliopoulos term on $S^1/Z_2$}

As an example consider a rigidly supersymmetric Abelian gauge theory coupled to massless matter fields in five dimensions, $D=4$, $d=1$. The orbifold is $S^1/Z_2$, where $S^1\cong\mathbb R/(2\pi R\mathbb Z)$, and the nontrivial $Z_2$ element acts on the circle as a reflection at the diameter. The fields are conveniently written in $D=4$ $N=1$ multiplets \cite{susy}: A 5D vector multiplet consists of a vector $A_M$, a real scalar $\Sigma$, a symplectic Majorana spinor represented as two Weyl spinors $\lambda_1,\lambda_2$, and three real auxiliary fields $\vec D$. It decomposes into a 4D $N=1$ vector multiplet $(A_\mu,\lambda_1,D)$, where $D\equiv D_3-\partial_5\Sigma$, and a chiral multiplet $((\Sigma+iA_5)/\sqrt{2},\lambda_2, D_1+iD_2)$. A 5D hypermultiplet decomposes into two chiral multiplets $H_+=(\phi_+,\psi_+,F_+)$, $H_-=(\phi_-,\psi_-,F_-)$; the subscript $\pm$ of the matter fields indicates their parities under the orbifold reflection, e.g.\ $\phi_\pm(x^1\ldots x^4,x^5)=\pm\phi_\pm(x^1\ldots x^4,-x^5)$. All fields are $2\pi$ periodic in the fifth coordinate because we have compactified on the circle. We take the gauge group to be Abelian, then the Fayet-Iliopoulos (FI) term
\be\label{tlfi} \mathcal L_{\mathrm{FI}}=\frac{1}{g^2}\xi(x^5)D \ee 
is allowed in the Lagrangian by both SUSY and gauge symmetry ($g$ is the $U(1)$ gauge coupling). It has been the subject of much recent investigation \cite{bccrs, gno, mp} how such an FI term can be generated at one-loop in the present model by radiative corrections, and what the consequences for phenomenology are.

The one-loop FI term is generated by scalar tadpole graphs with the $D$ auxiliary field. The relevant interaction term for a single 5D hypermultiplet of charge $q$, written in terms of the 4D component Lagrangian, is \cite{bccrs, gno}
\be\mathcal L_{D}=qD\left(|\phi_+|^2-|\phi_-|^2\right) \ee
In the worldline formalism, we can treat $D$ as one would treat an $x$-dependent mass term. Here we assume that there is no bulk mass term for the 
hypermultiplet. We comment on its relevance at the end of the paper.
The relevant contribution to the effective action in noncompact space for each of the scalars $\phi_+$ and $\phi_-$ is
\be\begin{split}\label{gammad} \Gamma_D=\int_0^\infty\frac{dT}{T}\,\int d^5 x_0\mspace{-24.0mu}\int\displaylimits_{y(0)=y(T)=0}\mspace{-24.0mu}\mathcal D^5y(\tau)\,\exp\left(-\int_0^T d\tau\,\left(\frac{1}{4}\dot y^2\pm q\,D(x_0+y)\right)\right) \end{split}\ee
To take the orbifolding into account, we replace the integral over closed loops in $\mathbb R^5$ by an integral over closed loops in $\mathbb R^4\times S^1/Z_2$, according to the prescription given in \eqref{toruspi},
\be\begin{split}\label{repl} \int d^{5} x_0\mspace{-24.0mu}\int\displaylimits_{y(0)=y(T)=0}\mspace{-24.0mu}\mathcal D^{5}y\,e^{-S_\mathrm{WL}[x_0+y]}\quad\rightarrow\quad&\frac{1}{2}\sum_{k\in\mathbb Z}\:\int d^4x_0\mspace{-24.0mu}\int\displaylimits_{y(0)=y(T)=0}\mspace{-24.0mu}\mathcal D^4y\mspace{-12.0mu}\int\displaylimits_{\substack{x^5(0)=x_0^5 \\ x^5(T)=x_0^5+2\pi Rk}}\mspace{-36.0mu}\mathcal Dx^5\,e^{-S_\mathrm{WL}[x_0+y,x^5]}\\
&\quad\pm\frac{1}{2}\sum_{k'\in\mathbb{Z}}\:\int d^4x_0\mspace{-24.0mu}\int\displaylimits_{y(0)=y(T)=0}\mspace{-24.0mu}\mathcal D^4y\mspace{-12.0mu}\int\displaylimits_{\substack{x^5(0)=x_0^5 \\ x^5(T)=-x_0^5+2\pi Rk'}}\mspace{-36.0mu}\mathcal Dx^5\,e^{-S_\mathrm{WL}[x_0+y,x^5]}\end{split}\ee
We can also split the path integral in the extra dimension into a base point integral and an integral over loops $\tilde y^5$ from $0$ to $0$ modulo orbifold symmetry. To do this, note that we can always write the 5-component of a given path $\tilde y(\tau)$ leading from $\tilde y^5(0)=0$ to $\tilde y^5(T)=2\pi Rk$ (or $\tilde y^5(T)=2\pi Rk-2x_0^5$) as a fixed path $z^5(\tau)$ between $0$ and $2\pi Rk$ (or $2\pi Rk-2x_0^5$), superimposed by some genuinely closed loop $y^5(\tau)$ with $y^5(0)=y^5(T)=0$. We choose $z^5(\tau)$ to be as simple as possible, interpolating linearly between $0$ and $2\pi Rk$ (I), or $0$ and $2\pi Rk-2x_0^5$ (II),
\begin{align*} z^5(\tau)&=2\pi Rk\frac{\tau}{T}\quad\textrm{(I)}, & z^5(\tau)&=2\left(\pi Rk -x_0^5\right)\frac{\tau}{T}\quad \textrm{(II)} \end{align*}
We thus end up with the following parameterizations for the two types of closed paths:
\begin{align*} x^5(\tau)&=x_0^5+2\pi Rk\frac{\tau}{T}+y^5(\tau)\quad\textrm{(I)}, & x^5(\tau)&=x_0^5\left(1-2\frac{\tau}{T}\right)+2\pi Rk\frac{\tau}{T}+y^5(\tau) \quad\textrm{(II)} \end{align*}%
so in the effective action we can make the following replacement for the free path integral
\be\begin{split}&\int d^5x_0\int \mathcal D^5y\exp\left(-\int_0^\infty d\tau \frac{1}{4}\dot y^2\right) \\&\rightarrow\quad \int d^4x_0 \int_0^{\pi R}dx_0^5\int \mathcal D^4 y\int \mathcal D y^5\,\exp\left(-\int_0^\infty d\tau\,\frac{1}{4}\dot y^2\right) \\
&\quad\qquad\times\frac{1}{2}\Biggl\{\sum_{k\in\mathbb Z}\exp\left(-\int_0^\infty d\tau\,\frac{1}{4}\left(\frac{2\pi Rk}{T}+\dot y^5\right)^2\right)\\
&\quad\qquad\qquad\pm\sum_{k'\in\mathbb Z}\exp\left(-\int_0^\infty d\tau\,\frac{1}{4}\left(\frac{2\pi Rk'}{T}+\dot y^5-\frac{2x_0^5}{T}\right)^2\right)\Biggr\} \end{split}\ee
with a $+$ ($-$) in the last line for even (odd) fields. Because in the last line only $\dot y^5$ depends on $\tau$, and the $\tau$ integral of $\dot y^5$ is zero, this simplifies to
\be\begin{split}\label{ugly}\ldots=& \int d^4x_0 \int_0^{\pi R}dx_0^5\int \mathcal D y\int \mathcal D y^5\,\exp\left(-\int_0^\infty d\tau\,\frac{1}{4}\left(\dot y^2+\left(\dot y^5\right)^2\right)\right)\\
&\quad\times\frac{1}{2}\Biggl\{\sum_{k\in\mathbb Z}\exp\left(-\frac{(\pi Rk)^2}{T}\right)\pm\sum_{k'\in\mathbb Z}\exp\left(-\frac{(\pi Rk'-x_0^5)^2}{T}\right)\Biggr\}\\
=&\frac{1}{2}\int d^4x_0 \int_0^{\pi R}dx_0^5\,(4\pi T)^{-5/2}\Biggl\{\sum_{k\in\mathbb Z}\exp\left(-\frac{(\pi Rk)^2}{T}\right)\pm\sum_{k'\in\mathbb Z}\exp\left(-\frac{(\pi Rk'-x_0^5)^2}{T}\right)\Biggr\} \end{split}\ee
Here we have performed the free path integrals according to \eqref{norm}.

To calculate the one-loop FI term, we add a coupling to $D$ as in \eqref{gammad} and consider the piece linear in $D$ in an expansion of the exponential:
\be\begin{split}\label{gammadpm}\Gamma_{D\pm}^{(1)}=&\pm \frac{q}{2}\int\frac{dT}{T}\,\int d^4x_0 \int_0^{\pi R}dx_0^5\,(4\pi T)^{-5/2}\int\mathcal D^4y\int\mathcal Dy^5\\
&\quad\times\Biggl\{\sum_{k\in\mathbb Z}\exp\left(-\frac{(\pi Rk)^2}{T}\right)\int_0^\infty d\tau\, D\left(x_0+y,\;x_0^5+y^5+2\pi Rk\frac{\tau}{T}\right)\\
&\quad\qquad\pm\sum_{k'\in\mathbb Z}\exp\left(-\frac{(\pi Rk'-x_0^5)^2}{T}\right)\int_0^\infty d\tau\,D\left(x_0+y,\;x_0^5+y^5+2(\pi Rk'-x_0^5)\frac{\tau}{T}\right)\Biggr\}\end{split}\ee
To evaluate this expression we expand $D$ around $(x_0,x_0^5)$,
\be D\left(x_0+y,\;x_0^5+y^5+2\pi Rk\frac{\tau}{T}\right)=\exp\left(y^\mu\partial_\mu+\left(y^5+2\pi Rk\frac{\tau}{T}\right)\partial_5\right)D(x_0,x_0^5)\ee
and perform a cumulant expansion writing
\be\begin{split}\label{cumexp}&\left< \exp\left(y^\mu\partial_\mu+\left(y^5+2\pi Rk\frac{\tau}{T}\right)\partial_5 \right)\right>\\
&\qquad=\exp\left(2\pi Rk\frac{\tau}{T}\partial_5+\frac{1}{2}\langle y(\tau)^2\rangle\partial^2+\frac{1}{2}\langle y^5(\tau)^2\rangle\partial_5^2+\ldots\right)\end{split}\ee
Accordingly,
\be\begin{split}\label{cumexp2}&\left<D\left(x_0+y,x_0^5+y^5+2\left(\pi Rk'-x_0^5\right)\frac{\tau}{T}\right)\right>\\
&\qquad=\exp\left(2(\pi Rk'-x_0^5)\frac{\tau}{T}\partial_5+\frac{1}{2}\langle y(\tau)^2\rangle\partial^2+\frac{1}{2}\langle y^5(\tau)^2\rangle\partial_5^2+\ldots\right)\,D(x_0,x_0^5)\end{split}\ee
The terms in the exponential series can now be calculated using the contraction rule \eqref{contr}, from which one in particular has $\langle y^\mu(\tau)y^\nu(\tau)\rangle=2 g^{\mu\nu}\tau(1-\tau/T)$. 

If we just consider the UV divergent parts, it turns out that in the UV limit $T\rightarrow 0$ only the leading term $D(x_0, x_0^5)$ causes divergences. To see why, note that only the $k'=0,1$ winding numbers can lead to infinities in that limit, as $x_0^5\in[0,2\pi]$. Hence we can write for the $k'$ sum
\be\begin{split}\int d^4x_0\int_0^{\pi R} dx_0^5\;\sum_{k'\in\mathbb{Z}}\exp\left(-\frac{(\pi Rk'-x_0^5)^2}{T}\right)&\\
\times\int_0^\infty d\tau\,\exp\Bigl(2(\pi Rk'-x_0^5)\frac{\tau}{T}\partial_5+&\frac{1}{2}\langle y(\tau)^2\rangle\partial^2+\frac{1}{2}\langle y^5(\tau)^2\rangle\partial_5^2+\ldots\Bigr)\,D(x_0,x_0^5) \\
=\int d^4x_0\int_0^{\pi R} dx_0^5\;\Biggl\{\exp\left(-\frac{(x_0^5)^2}{T}\right)\int_0^\infty d\tau\,&\exp\left(-2x_0^5\frac{\tau}{T}\partial_5+\frac{1}{2}\langle y^5(\tau)^2\rangle\partial_5^2\right)\\
+\left(x_0^5\leftrightarrow (x_0^5-\pi R)\right)&\Biggr\}\,D(x_0,x_0^5)+\ldots\\
=\int d^4x_0\int_0^{\pi R} dx_0^5\;\Biggl\{\exp\left(-\frac{(x_0^5)^2}{T}\right)\int_0^\infty d\tau\,&\Bigl[1-2x_0^5\frac{\tau}{T}\partial_5+\frac{1}{2}\langle y^5(\tau)^2\rangle\partial_5^2\\
&\qquad\qquad\qquad\qquad+\frac{1}{2}\left(-2x_0^5\frac{\tau}{T}\partial_5\right)^2+\ldots\Bigr]\\
+\left(x_0^5\leftrightarrow (x_0^5-\pi R)\right)&\Biggr\}\,D(x_0,x_0^5)+\ldots
\end{split}\ee
where we have kept explicit only potentially UV-divergent terms. However, the contributions from the derivative terms cancel for an arbitrary even $D$ as $T\rightarrow 0$, so only the constant term gives rise to infinities. Furthermore, the $k$ sum part of \eqref{gammadpm} cancels between odd and even fields, since each even scalar is paired up with an odd scalar in our model, so we get the contribution to the overall one-loop FI term from a single hypermultiplet by adding an odd and an even part. We are thus left with
\be\begin{split}\label{gammapm}\Gamma_{D+}^{(1)}+\Gamma_{D-}^{(1)}=&q\int\frac{dT}{T}\,\int d^4x_0 \int_0^{\pi R}dx_0^5\,(4\pi T)^{-5/2}\\
&\qquad\qquad\times\Biggl\{\exp\left(-\frac{(x_0^5)^2}{T}\right)+\exp\left(-\frac{(\pi R-x_0^5)^2}{T}\right)\Biggr\}\,D(x_0, x_0^5)\,T\\
&+\textrm{finite}\end{split}\ee
Retaining the dependence on the extra dimension, this gives for the one-loop FI coefficient $\xi^{(1)}(x_0^5)$ by comparison with the tree-level coefficient in \eqref{tlfi}
\be\label{xione}\xi^{(1)}(x_0^5)=qg^2\int_0^\infty\frac{dT}{T}\, (4\pi T)^{-5/2}\left\{\exp\left(-\frac{(x_0^5)^2}{T}\right)+\exp\left(-\frac{(\pi R-x_0^5)^2}{T}\right)\right\} T +\textrm{finite}\ee
For $x_0^5$ in the bulk the integral is also UV finite: we get a divergent part only on the branes. Cutting off the proper time integral at its lower bound at $T_0=1/\Lambda^2$, we find by convolution with a test function
\be\label{susyfi}\xi^{(1)}(x_0^5)=\frac{qg^2}{32\pi^2}\left(\left(\delta(x_0^5)+\delta(x_0^5-\pi R)\right)\Lambda^2+\frac{1}{4}\left(\delta''(x_0^5)+\delta''(x_0^5-\pi R)\right)\log\Lambda^2\right)+\textrm{finite} \ee
in accordance with \cite{bccrs, gno, mp}. In a framework of local 
supersymmetry, the $\Lambda^2$ term breaks gauge invariance. In fact, 
with several hypermultiplets, the absence of mixed gravitational-$U(1)$ anomalies 
requires that $\sum q_i=0$. With this condition the divergent
piece vanishes identically.

\section{Bulk-brane couplings}

In this section we consider bulk fields with brane-localized operators, by means of an example similar to the one above. As before we take SUSY QED on $S^1/Z_2$, but include SUSY breaking masses for the scalar fields located on one of the orbifold boundaries. We will see that this may induce a one-loop FI term even if the usual condition for its absence in supersymmetric models is satisfied --- namely, that the charges sum up to zero between the hypermultiplets.

Assume the Lagrangian contains a soft SUSY breaking term localized 
on the $x^5=0$ brane of the form
\be\mathcal L_{m^2}=Rm^2\,\delta(x^5)\,|\phi_+|^2 
\label{softmass}
\ee
(brane couplings to odd fields would make little sense since odd fields have no support on the boundaries). The worldline action for an even scalar then reads
\be S_\mathrm{WL}=\int_0^T\left(\frac{1}{4}\dot x^2+Rm^2\,\delta\left(x^5(\tau)\right)+qD\left(x(\tau)\right)+\ldots\right)\ee
Exponentiating this action requires us to define what we mean by arbitrary powers of the delta function: If a worldline is to give a contribution to the effective action involving $n$ powers of $Rm^2$, it has to pass through the $x^5=0$ brane $n$ times at some $\tau_0^1\ldots\tau_0^n$ with $0<\tau_0^1<\ldots<\tau_0^n<T$. The integration over all closed paths leading from $x_0^5$ to $x_0^5$ passing $n$ times through the brane thus becomes a proper time ordered product of $n$ proper time integrals, with the integrand consisting of $n+1$ path integrals over paths of length $(\tau_0^{i+1}-\tau_0^{i})$ (where $0\leq i\leq n$), connecting $0$ with $0$ (or $x_0^5$ with $0$ for the first, setting $\tau_0^0\equiv 0$, or $0$ with $x_0^5$ for the last, setting $\tau_0^{n+1}\equiv T$):
$$ \int_0^Td\tau_0^n\int_0^{\tau_0^n}d\tau_0^{n-1}\:\cdots\:\int_0^{\tau_0^2}d\tau_0^1\,\int\displaylimits_{\substack{y^5_0(0)=x_0^5 \\ y^5_0(\tau_0^1)=0}}\mspace{-18.0mu}\mathcal Dy^5_0\int\displaylimits_{\substack{y^5_1(0)=0 \\ y^5_1(\tau_0^2-\tau_0^1)=0}}\mspace{-24.0mu}\mathcal Dy^5_1\quad\cdots\:\mspace{-24.0mu}\int\displaylimits_{\substack{y^5_n(0)=0 \\ y^5_n(T-\tau_0^n)=x_0^5}}\mspace{-24.0mu}\mathcal Dy^5_n$$
On the orbifold, we also have to sum over all paths connecting points that are identified with the respective start and end points.

Let us illustrate the procedure above by calculating the contribution to first order in $Rm^2$ to the one-loop FI term. If a path is to give a contribution $\sim Rm^2$, it has to pass through the brane once at some $\tau_0$ with $0<\tau_0<T$, hence we have an integration over all $\tau_0$, an integration over all paths leading from $x_0^5$ to $0$, and an integration over all paths leading from $0$ back to $x_0^5$. Furthermore, we identify $x_0^5\sim \pm x_0^5+2\pi kR$ and $0\sim 2\pi kR$ under the orbifold symmetry. 

Consider the 5-component of a given path $x^5(\tau)$ which leads from $x_0^5$ to $0$ (or from $0$ to $x_0^5$) modulo orbifold symmetry. As above, it can be represented by a linearly parametrized fixed path $z^5(\tau)$ and a closed loop $y^5(\tau)$. We thus have the following parametrizations:
\be\begin{split} x^5_\mathrm{Ia}(\tau)&= x^5_\mathrm{IIa}(\tau)=
2\pi Rk\frac{\tau}{\tau_0} +x_0^5\frac{\tau_0-\tau}{\tau_0}+y^5_a(\tau) \\
 x^5_\mathrm{Ib}(\tau)&=2\pi Rk\frac{T-\tau}{T-\tau_0} +(x_0^5+2\pi R\tilde k)\frac{\tau-\tau_0}{T-\tau_0}+y^5_b(\tau)\\
 x^5_\mathrm{IIb}(\tau)&=2\pi Rk\frac{T-\tau}{T-\tau_0}+ (-x_0^5+2\pi R\tilde k)\frac{\tau_0-\tau}{T-\tau_0}+y^5_b(\tau)
\end{split}\ee
with $y^5_a(0)=y^5_a(\tau_0)=y^5_b(\tau_0)=y^5_b(T)=0$. (Ia) and (IIa) lead from \mbox{$x^5(0)=x_0^5$} to $x^5(\tau_0)=0+2\pi Rk$. (Ib) and (IIb) lead back from $x^5(\tau_0)=2\pi Rk$ to $x^5(T)=\pm x_0^5+2\pi R\tilde k$. 

Now we are ready to write down the one-loop effective action to first order in $Rm^2$. We make the following replacement with respect to the massless path integral in uncompactified space, eq.\ \eqref{gammad},
\be\label{bigeq}\begin{split} \int& d^4x_0\int_{-\infty}^{\infty} dx_0^5\int\mathcal D^4y\int\mathcal Dy^5 \exp\left(-\int_0^T\frac{1}{4}\left(\dot y^2+\left(\dot y^5\right)^2\right)\right) \\
&\rightarrow \,-Rm^2\int d^4 x_0\int_0^{\pi R} dx_0^5\int\mathcal D^4y\,\exp\left(-\int_0^T\frac{1}{4}\dot y^2\right)\int_0^T d\tau_0 \\
&\quad\times\Biggl\{\frac{1}{2}\int \mathcal D y^5_a\sum_{k\in\mathbb{Z}}\exp\left(-\int_0^{\tau_0} d\tau\,\frac{1}{4}\left(\dot x^5_\mathrm{Ia}\right)^2\right)\,\int\mathcal D y^5_b \sum_{\tilde k\in\mathbb{Z}}\exp\left(-\int_{\tau_0}^{T} d\tau\,\frac{1}{4}\left(\dot x_\mathrm{Ib}^5\right)^2\right)\\
&\quad\qquad +
\left({\rm I}\leftrightarrow {\rm II}\right)\Biggr\} 
\end{split}\ee
After shifting some of the sums setting $k'=\pm k\mp\tilde k$, performing the free path integrals, and finally substituting $\tau_0-T/2\equiv\tau$, this becomes
\be \begin{split}\ldots=&-Rm^2\int d^4x_0\int_{0}^{\pi R} dx_0^5\,\left(4\pi\right)^{-3} T^{-2}\int_0^T d\tau_0\,\left(\tau_0(T-\tau_0)\right)^{-1/2} \\
&\times\sum_{k\in\mathbb{Z}}\exp\left(-\frac{1}{4}\frac{(2\pi Rk-x_0^5)^2}{\tau_0}\right)\sum_{k'\in\mathbb{Z}}\exp\left(-\frac{1}{4}\frac{(2\pi Rk'+x_0^5)^2}{T-\tau_0}\right)\\
=&-\frac{Rm^2}{(4\pi)^3}\int d^4x_0\int_{0}^{\pi R} dx_0^5\,T^{-2}\int_{-T/2}^{T/2} d\tau\,\left(T^2/4-\tau^2\right)^{-1/2} \\
&\times\sum_{k,k'\in\mathbb{Z}}\exp\left(-\frac{1}{4}\frac{(T/2-\tau)(2\pi Rk-x_0^5)^2+(T/2+\tau)(2\pi Rk'-x_0^5)^2}{T^2/4-\tau^2}\right)
\end{split}\ee
The part of the effective action relevant for the renormalization of the FI term is then once more obtained by adding a mass-term like coupling to $D$ and considering just the linear part. The terms in an expansion like \eqref{cumexp} can be calculated to any order; here we again restrict ourselves to the divergent part. The effective action then reads, up to finite terms which we dropped when setting $D(x,x^5)\approx D(x_0,x_0^5)$, and conveniently substituting $\tau=T/2\sin\tilde\tau$,
\be\begin{split}\Gamma^{(1)}=&\Gamma^{(1)}_{D+}+\Gamma^{(1)}_{D-}\,-\,\frac{qRm^2}{(4\pi)^3}\int_0^{\infty}\frac{dT}{T}\int d^4x_0\int_{0}^{\pi R} dx_0^5\, T^{-2}\int_{-\pi/2}^{\pi/2} d\tilde\tau\\
&\times \sum_{k,k'\in\mathbb{Z}}
\exp\left( -\frac{(1-\sin\tilde\tau)(2\pi Rk-x_0^5)^2+
(1+\sin\tilde\tau)(2\pi Rk'-x_0^5)^2}{2T\cos^2\tilde\tau }\right)\,D(x_0,x_0^5)\,T \end{split}\ee
with $\Gamma^{(1)}_{D+}+\Gamma^{(1)}_{D-}$ as in eq.\ \eqref{gammapm}. From this expression it is clear that the only divergences in the additional term come from $k=k'=0$ and $x_0^5=0$ --- it is of course not surprising to find the FI term localized on the same brane as the SUSY breaking mass. Keeping again the profile in the extra dimension, we obtain a divergent contribution to the one-loop FI term of
\be\begin{split} \xi^{(1)}_{Rm^2}(x_0^5)&=-\frac{g^2q Rm^2}{32\pi^{5/2}}\int_{1/\Lambda^2}^\infty dT\, T^{-3/2}\,\delta(x_0^5)\\
&=-\frac{g^2qRm^2}{16\pi^{5/2}}\Lambda\,\delta(x_0^5)\end{split} 
\label{fi41}
\ee
(again with a cutoff at $T_0=1/\Lambda^2$).
This new result deserves a comment. If one introduces an odd bulk mass
$\overline{M}(x^5)=M\epsilon (x^5)$  for the hypermultiplet, then the 
masses of the bosonic components get an additional contribution 
\be
2M[\delta (x^5)-\delta (x^5-\pi R)](|\phi_{+}|^2-|\phi_{-}|^2)~.
\label{bulkM}
\ee 
As was shown in \cite{bccrs}, this term creates a linear divergence
in the FI term. As we have seen the
soft SUSY breaking boundary term (\ref{softmass}), which has a form similar 
to (\ref{bulkM})
with the replacement $m^2\to 2M/R$, also induces such a divergent 
contribution into  the FI term. Thus it acts as an explicitly orbifold 
parity violating bulk mass operator.

It is straightforward to compute also the $\mathcal O\left((Rm^2)^2\right)$ contribution to the one-loop renormalization of the FI term (cubic and higher order terms in $Rm^2$ are finite by power counting): split the paths passing the $x^5=0$ brane twice into 3 parts, choose the parametrizations, and do the integrals. The result is
\be\xi^{(1)}_{(Rm^2)^2}(x_0^5)=
\frac{g^2q(Rm^2)^2}{64\pi^3}\log\Lambda^2\,\delta(x_0^5)
\label{fi42}
\ee
Up to finite terms the total one-loop FI coefficient is thus given by
\be\xi^{(1)}(x_0^5)=\xi^{(1)}_0(x_0^5)+\xi^{(1)}_{Rm^2}(x_0^5)+\xi^{(1)}_{(Rm^2)^2}(x_0^5)\ee
with $\xi^{(1)}_0(x_0^5)$ as in unbroken SUSY, equation \eqref{susyfi}.
In presence of several charged hypermultiplets, for a universal soft mass, the 
condition $\Sigma q_i=0$ ensures that the contributions (\ref{fi41}),
(\ref{fi42}) vanish. However, with non-universal soft masses, 
such contributions do arise and  are relevant for the FI term corresponding 
to the $U(1)_Y$ hypercharge gauge factor of the Standard Model.

\section{Conclusions}

In this paper we have presented an alternative method to perform perturbative calculations in orbifold field theory, by adapting the well-known worldline formalism to orbifolds. We have shown that, with the worldline formalism relying on path integration over closed worldlines, its use in orbifolds requires the inclusion of paths that are closed only modulo orbifold symmetry in the path 
integral. 
The main advantages of this procedure are that the singular structure in the extra dimension becomes quite transparent, and that a universal cutoff in the proper time parameter $T$ regulates also the 4D part.
We have applied this method to reproduce the one-loop Fayet-Iliopoulos term in supersymmetric QED on the $S^1/Z_2$ orbifold. It has turned out that it is also possible to treat brane-localized operators for bulk fields in this formalism, by appropriately splitting the corresponding worldlines. As a sample calculation, the one-loop renormalization of the Fayet-Iliopoulos term was calculated in the presence of soft SUSY breaking scalar masses on one of the branes.

\vspace{0.5cm}

\hspace{-0.7cm}{\bf Acknowledgments}

\vspace{0.2cm} 
\hspace{-0.7cm}We would like to thank F.~Paccetti Correia for very useful 
discussions and M.~Olechowski for some remarks which helped to correct a previous version of this paper.

\end{document}